# Preliminary Estimators of Population Mean using Ranked Set Sampling in the Presence of Measurement Error and Non-Response Error

*Rajesh Singh*[1] and **Anamika Kumari**[2*]

Department of Statistics, Institute of Science, Banaras Hindu University, Varanasi

rsinghstat@gmail.com

*anamikatiwari1410@gmail.com (corresponding author)

**Abstract**

In order to estimate the population mean in the presence of both non-response and measurement errors that are uncorrelated, the paper presents some novel estimators employing ranked set sampling by utilizing auxiliary information. Up to the first order of approximation, the equations for the bias and mean squared error of the suggested estimators are produced, and it is found that the proposed estimators outperform the other existing estimators analysed in this study. Investigations using simulation studies and numerical examples show how well the suggested estimators perform in the presence of measurement and non-response errors. The relative efficiency of the suggested estimators compared to the existing estimators has been expressed as a percentage, and the impact of measurement errors has been expressed as a percentage computation of measurement errors.

## 1. INTRODUCTION

Sampling is important because of many reasons like cost and time constraints. Auxiliary information is additional information utilized to improve the efficiency of the estimator. The use of auxiliary information can be done at various stages. Highly correlated auxiliary information is usually well known if not available then it might have been gathered from earlier surveys. In this context, good examples of estimation techniques are ratio, product, and regression.

In sampling, there is a constant desire for enhancement of results covering efficiencies of the estimator, cost, complications, and time. Ranked Set Sampling (RSS) is an improved sampling method over Simple Random Sampling (SRS). In a variety of disciplines, including medical, farming related sciences, earth sciences, and many fields of statistics and mathematics, RSS is more affordable than SRS. McIntyre (1952) was the first to explain RSS technique for estimation of the population mean. Takahashi and Wakimoto (1968) gave the necessary mathematical theory of RSS. When considering the cases of perfect and imperfect ranking, the mean under RSS has been shown to be an unbiased estimate of the population mean by Dell and Clutter (1972).

While conducting sampling survey, we usually come across non-sampling errors like measurement error (ME) and non-response error (NRE). In a sampling survey, it is believed that the observed values are true when estimating the population parameters. We never came across this ideality accounting for errors in measurement. The gap between observed values and their corresponding true values is referred to as the error. A respondent may purposefully or unintentionally report their income in a household survey differently (more or less) than their actual income. Shalabh (1997) used the ratio method for estimation in presence of ME. Singh and Karpe (2001) and Kumar et al. (2011) proposed a ratio-product estimator and some ratio-type estimators respectively for finite population mean under MEs. Several authors have examined the issue of estimating the finite population mean under measurement error using auxiliary information, including Malik and Singh (2013), Singh et al. (2014), Khalil et al. (2018), Zahid & Shabbir (2018), and Singh et al. (2019).

Vishwakarma & Singh (2022) have proposed ratio, product, difference, and exponential estimators in the presence of ME using RSS.

Many sampling surveys employ the mail questionnaire to collect information due to financial restrictions. Non-response in sample surveys is a widespread issue that affects mail surveys more than in-person interviews. Non-response is the failure to collect data from a few units of the population that was selected for the purpose of the study. The first researchers that investigated the non-response problem was Hansen & Hurwitz in 1946. They suggested a sampling strategy that comprises enumerating the subsample through personal interviews after taking a subsample of non-respondents from the initial mail attempt. El-Badry (1956) extended the method of Hansen & Hurwitz.

Authors such as Cochran (1977), Khare and Srivastava (1993), Singh et al. (2009) have studied the problem of non-response. Bouza and Harrera (2013) have considered problem of the non-response under RSS. For recent work in RSS you can refer Shabbir (2022).

The issue of estimation employing the RSS technique in the context of errors (ME and NRE) is not given much attention. In this paper, our aim is to study estimators that may enhance true estimation of population mean under RSS when there are presence of errors (ME and NRE) simultaneously on both the study and auxiliary variables.

In search of efficient estimators, we proposed some new estimators of population mean under RSS when there are errors (ME and NRE). These new estimators are expected to give a more precise and efficient estimate of the population mean than the existing estimators considered in this paper.

## 2. Sampling Methodology

In ranked set sampling (RSS), we rank randomly selected units from the population merely by observation or prior experience after which only a few of these sampled units are measured. In RSS, m independent random sets, each of size m, are selected from the population. Each unit in the set has an equal chance of being chosen. Each random set's constituents are ranked according to the auxiliary variable's characteristic. Next, the smallest unit in the first ordered set is chosen, then the next-smallest unit in the second ordered set is chosen. This process is carried out in this manner until the largest rank in the $m^{th}$ set is chosen. rm (= n) units have been measured throughout this process as this cycle is repeated r times.

Consider a finite population U = $(U_1, U_2...U_N)$ based on N identifiable units with a study variable Y and auxiliary variables X. Using the RSS technique we extract a sample of size n=rm units from it. Let $(x_{mej\,(l)}, y_{mej\,[l]})$ l=1, 2....m, j=1, 2...r be observed values on X and Y corresponding to true values $(X_{j(l)}, Y_{j[l]})$ l=1, 2....m, j=1, 2...r respectively of the sets of the $l^{th}$ units in the $j^{th}$ cycle. Let $u_{j[l]} = y_{mej\,[l]} - Y_{j[l]}$ and $v_{j(l)} = x_{mej\,(l)} - X_{j(l)}$ be the measurement errors on the study and auxiliary variable respectively. The error terms (u, v) follow the normal distribution, which has a mean of 0 and a variance of $(\sigma_u^2, \sigma_v^2)$, and also these error terms are independent of both variables (X, Y). Let $\rho_{uv}$ represent the correlation coefficient between the errors (u, v) in the case of uncorrelated ME it is zero, and also Y and X are correlated with $\rho_{xy}$.

Let the unbiased estimators of population means $\bar{Y}, \bar{X}$ be
$\bar{y}_{me} = \frac{1}{n}\sum_{i=1}^{n} y_{me\,[i]} = \frac{1}{rm}\sum_{l=1}^{k}\sum_{j=1}^{r} y_{mj\,[l]}$, $\bar{x}_{me} = \frac{1}{n}\sum_{i=1}^{n} x_{me\,(i)} = \frac{1}{rm}\sum_{l=1}^{k}\sum_{j=1}^{r} x_{mj\,(l)}$,

for the study and auxiliary variables, but when it comes to variance

$E(s_{mey}^2) = \sigma_y^2 + \sigma_u^2 = \frac{1}{N-1}\sum_{i=1}^{N}(Y_{[i]} - \bar{Y})^2 + \frac{1}{N-1}\sum_{i=1}^{N}(U_{[i]} - \bar{U})^2$ and

$(s_{mex}^2) = \sigma_x^2 + \sigma_v^2 = \frac{1}{N-1}\sum_{i=1}^{N}(X_{(i)} - \bar{X})^2 + \frac{1}{N-1}\sum_{i=1}^{N}(V_{(i)} - \bar{V})^2$.

Here $s_{mey}^2 = \frac{1}{n-1}\sum_{i=1}^{n}(y_{me\,[i]} - \bar{y}_{me})^2$, $s_{mex}^2 = \frac{1}{n-1}\sum_{i=1}^{n}(x_{me\,(i)} - \bar{x}_{me})^2$.

According to Hansen & Hurwitz (1946) method, from a finite population of size N, we use the SRSWOR method to generate a sample S of size n. Let $n_1$ units respond the survey on the first try, whereas $n_2 (= n - n_1)$ units fail to do so. A portion of the non-responding units ($n_2' = n_2/k; k > 1$) is included in the sample as a result of further efforts made to contact them. As a result, we end up with a sample that of size $n = n_1 + n_2'$. This makes it possible to divide the total population into two complimentary categories known as response and non-response groups. Let $(Y_{ji}, X_{ji})$; i = 1, 2, ..$N_j$ ; j = 1, 2 be population units of the study variable (Y) and the auxiliary variable (X) in the two groups. When there is non-response on Y, Hansen and Hurwitz (1946) recommended the following unbiased estimator:

$$\bar{y}_{srs}^* = w_1\bar{y}_1 + w_2\bar{y}_2' \qquad (2.1)$$

The variance of $\bar{y}_{srs}^*$ is shown by

$$Var(\bar{y}_{srs}^*) = \left(\frac{1}{n} - \frac{1}{N}\right)\sigma_y^2 + \frac{W_2(k-1)}{n}\sigma_{y2}^2 \qquad (2.2)$$

where,

$$\bar{y}_1 = \frac{1}{n_1}\sum_{i=1}^{n_1} y_{1i}, \bar{y}_2' = \frac{1}{n_2'}\sum_{i=1}^{n_2'} y_{2i}, w_j = \frac{n_j}{n}; j = 1,2$$

and

$$\bar{Y}_1 = \frac{1}{N_1}\sum_{i=1}^{N_1} Y_{1i}, \quad \bar{Y}_2' = \frac{1}{N_2}\sum_{i=1}^{N_2} y_{2i}, \quad \bar{Y} = \frac{1}{N}\sum_{i=1}^{N} Y_i = W_1\bar{Y}_1 + W_2\bar{Y}_2$$

$$\sigma_y^2 = \frac{1}{N-1}\sum_{i=1}^{N}(Y_i - \bar{Y})^2, \quad \sigma_{y2}^2 = \frac{1}{N_2-1}\sum_{i=1}^{N_2}(Y_{2i} - \bar{Y})^2, \quad W_j = \frac{N_j}{N}; j = 1,2$$

An auxiliary variable X can yield similar results.

$$Cov(\bar{y}_{srs}^*, \bar{x}_{srs}^*) = \left(\frac{1}{n} - \frac{1}{N}\right)\sigma_{yx} + \frac{W_2(k-1)}{n}\sigma_{yx2} \tag{2.3}$$

In this paper, we gathered a sample from both groups (respondent and non-respondent) by using RSS.

$$\bar{y}_{rss}^* = w_1 \bar{y}_{me\,1,rss} + w_2 \bar{y}_{me\,2,rss}' \tag{2.4}$$

where, $\bar{y}_{me\,1,rss}$ is the sample mean based on $n_1$ units acquired at first attempt, while $\bar{y}_{me\,2,rss}'$ is the sample mean calculated on the basis of $n_2'$ units acquired at second attempt. $\bar{y}_{rss}^*$ is also an unbiased estimator, the variance of $\bar{y}_{rss}^*$ given by

$$Var(\bar{y}_{rss}^*) = \eta\sigma_y^2 - D_y^2 + w_2(k-1)\big(\eta\sigma_{y2}^2 - D_{y2}^2\big) + \eta\sigma_v^2 - D_v^2 \\ + w_2(k-1)(\eta\sigma_{v2}^2 - D_{v2}^2) \tag{2.5}$$

For the auxiliary variable, similar formulas can be constructed as follows:

$$Var(\bar{x}_{rss}^*) = \eta\sigma_x^2 - D_x^2 + w_2(k-1)(\eta\sigma_{x2}^2 - D_{x2}^2) + \eta\sigma_u^2 - D_u^2 \\ + w_2(k-1)(\eta\sigma_{u2}^2 - D_{u2}^2) \tag{2.6}$$

$$Cov(\bar{y}_{rss}^*, \bar{x}_{rss}^*) = \eta\sigma_{yx} - D_{yx} + w_2(k-1)\big(\eta\sigma_{yx2} - D_{yx2}\big) \tag{2.7}$$

where,

$D_y^2 = \frac{1}{m^2 r}\sum_{i=1}^{k}(\mu_{[iy]} - \bar{Y})^2$,

$D_x^2 = \frac{1}{m^2 r}\sum_{i=1}^{k}(\mu_{(ix)} - \bar{X})^2$,

$D_{yx} = \frac{1}{m^2 r}\sum_{i=1}^{k}(\mu_{[iy]} - \bar{Y})(\mu_{(ix)} - \bar{X})$,

$D_{y2}^2 = \frac{1}{m^2 r_2'}\sum_{i=1}^{k}(\mu_{[iy2]} - \bar{Y})^2$,

$D_{x2}^2 = \frac{1}{m^2 r_2'}\sum_{i=1}^{k}(\mu_{(ix2)} - \bar{X})^2$,

$D_{yx2} = \frac{1}{m^2 r_2'}\sum_{i=1}^{k}(\mu_{[iy2]} - \bar{Y})(\mu_{(ix2)} - \bar{X})$,

$D_u^2 = \frac{1}{m^2 r} \sum_{i=1}^{k} (\mu_{[iu]} - \bar{U})^2,$

$D_v^2 = \frac{1}{m^2 r} \sum_{i=1}^{k} (\mu_{(iv)} - \bar{V})^2,$

$D_{u2}^2 = \frac{1}{m^2 r_2'} \sum_{i=1}^{k} (\mu_{[iu2]} - \bar{U})^2,$

$D_{v2}^2 = \frac{1}{m^2 r_2'} \sum_{i=1}^{k} (\mu_{(iv2)} - \bar{V})^2,$

$\eta = \frac{1}{mr}.$

where $\mu_{[iy]}$ and $\mu_{(ix)}$ are the means of the $i^{th}$ ranked set and are given by

$\mu_{[iy]} = \frac{1}{r} \sum_{l=1}^{r} y_{i(i)l} \, , \mu_{(ix)} = \frac{1}{r} \sum_{l=1}^{r} x_{i(i)l}.$

Keep in mind that various notations are employed and the set size m is maintained constant.

$n_1 = mr_1, n_2 = mr_2, n_2' = mr_2', r = r_1 + r_2', n = mr, k = \frac{n_2}{n_2'} = \frac{r_2}{r_2'}.$

To obtain the bias and MSE of the estimators, we write

$\bar{y}_{rss}^* = \bar{Y}(1 + \epsilon_0),$
$\bar{x}_{rss}^* = \bar{X}(1 + \epsilon_1).$

E ($\epsilon_0$)=E ($\epsilon_1$)=0,

$E(\epsilon_0^2) = \frac{1}{\bar{Y}^2} [\eta \sigma_y^2 - D_y^2 + w_2(k-1)(\eta \sigma_{y2}^2 - D_{y2}^2) + \eta \sigma_v^2 - D_v^2$
$\qquad + w_2(k-1)(\eta \sigma_{v2}^2 - D_{v2}^2)] = V_y$

$E(\epsilon_1^2) = \frac{1}{\bar{X}^2} [\eta \sigma_x^2 - D_x^2 + w_2(k-1)(\eta \sigma_{x2}^2 - D_{x2}^2) + \eta \sigma_u^2 - D_u^2$
$\qquad + w_2(k-1)(\eta \sigma_{u2}^2 - D_{u2}^2)] = V_x$

$E(\epsilon_0 \epsilon_1) = \frac{1}{\bar{Y}\bar{X}} [\eta \sigma_{yx} - D_{yx} + w_2(k-1)(\eta \sigma_{yx2} - D_{yx2})] = V_{yx}$

### 3. Existing Estimators

The usual unbiased estimator for the population mean $\bar{Y}$ in the presence of errors using RSS technique is given by

$$\bar{y}_{rss}^* = w_1 \bar{y}_{me1,rss} + w_2 \bar{y}_{me2,rss}' \qquad (3.1)$$

The variance of the estimator $\bar{y}_{rss}^*$ is given by

$$Var(\bar{y}_{rss}^*) = \bar{Y}^2 V_y \qquad (3.2)$$

The ratio estimator under RSS for the population mean $\bar{Y}$ in the presence of errors

$$\bar{y}_{Re} = \bar{y}_{rss}^* \frac{\bar{X}}{\bar{x}_{rss}^*} \qquad (3.3)$$

The MSE of the estimator $\bar{y}_{Re}$ is shown by

$$MSE(\bar{y}_{Rme}) = \bar{Y}^2(V_y + V_x - 2V_{yx}) \quad (3.4)$$

The regression estimator under RSS for the population mean $\bar{Y}$ in the presence of errors

$$\bar{y}_{De} = \bar{y}_{rss}^* + \beta(\bar{X} - \bar{x}_{rss}^*) \quad (3.5)$$

The MSE of the estimator $\bar{y}_{De}$ is shown by

$$MSE(\bar{y}_{De}) = \bar{Y}^2\left(V_y - \frac{V_{yx}^2}{V_x}\right) \quad (3.6)$$

The exponential estimator under RSS for the population mean $\bar{Y}$ in the presence of errors is given by

$$\bar{y}_{exp} = \bar{y}_{me} \exp\left(\frac{\bar{X} - \bar{x}_{rss}^*}{\bar{X} + \bar{x}_{rss}^*}\right) \quad (3.7)$$

The MSE of the estimator $\bar{y}_{exp}$ is shown by

$$MSE(\bar{y}_{exp}) = \bar{Y}^2\left(V_y + \frac{V_x}{4} - V_{yx}\right) \quad (3.8)$$

## 4. Proposed Estimators

There isn't one estimator that works well in every circumstance. Therefore, having estimators that provide minimum MSE and high precision are preferable. The goal of this section is to create estimators that operate effectively over a wider domain. We adopted Mishra et al. (2017) estimator under RSS in the presence of errors (ME and NRE) and also proposed two new estimators of finite population mean under non-response error and measurement error by utilizing auxiliary information.

1.) $P_1 = \bar{y}_{rss}^*(g_1 + 1) + g_2 \log\left(\frac{\bar{x}_{rss}^*}{\bar{X}}\right)$ \quad (4.1)

where the constants $g_1$ and $g_2$ ensure that the estimators' MSE is kept to a minimal.

Expressing the estimator $P_1$ given in equation (4.1) in terms of $\epsilon$'s we get

$$P_1 = \bar{Y}(1 + \epsilon_0)(g_1 + 1) + g_2 \log\left(\frac{\bar{X}(1 + \epsilon_1)}{\bar{X}}\right) \quad (4.2)$$

Taking expectations up to the first order approximation, we get mean square error (MSE),

$$MSE(P_1) = \bar{Y}^2 V_y + g_1^2 A_1 + g_2^2 B_1 + 2g_1 C_1 + 2g_2 D_1 + 2g_1 g_2 E_1 \quad (4.3)$$

where,

$$A_1 = \bar{Y}^2(1 + V_y)$$

$$B_1 = V_x$$

$$C_1 = \bar{Y}^2 V_y$$

$$D_1 = \bar{Y} V_{yx}$$

$$E_1 = \bar{Y}\left(V_{yx} - \frac{1}{2}V_x\right)$$

To find out the minimum MSE for $P_1$, we partially differentiate equation (4.3) wrt $g_1$ & $g_2$ and equating to zero we get

$$g_1^* = \frac{B_1 C_1 - D_1 E_1}{E_1^2 - A_1 B_1} \tag{4.4}$$

$$g_2^* = \frac{A_1 D_1 - C_1 E_1}{E_1^2 - A_1 B_1} \tag{4.5}$$

Putting the optimum value of $g_1$ & $g_2$ in the equation (4.3), we get a minimum value of MSE of $P_1$ as

$$\text{Min } MSE = C_1 + \frac{B_1 C_1^2 + A_1 D_1^2 - 2C_1 D_1 E_1}{E_1^2 - A_1 B_1} \tag{4.6}$$

2.) $P_2 = g_3 \bar{y}_{rss}^* + g_4 \exp\left(\frac{\bar{X} - \bar{x}_{rss}^*}{\bar{X} + \bar{x}_{rss}^*}\right)\left(1 + \log \frac{\bar{x}_{rss}^*}{\bar{X}}\right)$ (4.7)

Expressing $P_2$ given in equation (4.7) in terms of $\epsilon$'s we get

$$P_2 = g_3 \bar{Y}(1 + \epsilon_0) + g_4 \exp\left(\frac{-\epsilon_1}{2 + \epsilon_1}\right)(1 + \log(1 + \epsilon_1)) \tag{4.8}$$

$$P_2 - \bar{Y} = (g_3 - 1)\bar{Y} + g_3 \bar{Y} \epsilon_0 + g_4 \left(1 + \frac{\epsilon_1}{2} - \frac{5\epsilon_1^2}{8}\right) \tag{4.9}$$

Bias $(P_2) = \bar{Y}(g_3 - 1) + g_4 \left[1 - \frac{5}{8} V_x\right]$ (4.10)

**CASE 1: SUM OF WEIGHTS IS UNITY ($g_3 + g_4 = 1$)**

The MSE of the estimator $P_2$ is shown as

$MSE(P_2) = \bar{Y}^2 [V_y + g_4^2 V_x - 2g_4 V_{yx}]$ (4.11)

To find out the minimum MSE for $P_2$, we partially differentiate equation (4.11) wrt $g_4$, and equating to zero we get

$$g_4^* = \frac{V_{yx}}{V_x} \tag{4.12}$$

Putting the optimum value of $g_4$ in the equation (4.11), we get a minimum MSE of $P_2$ as

Min MSE = $\bar{Y}^2 \left(V_y - \frac{V_{yx}^2}{V_x}\right)$ (4.13)

## CASE 2: THE SUM OF WEIGHTS IS FLEXIBLE ($g_3 + g_4 \neq 1$)

$$P_2 - \bar{Y} = (g_3 - 1)\bar{Y} + g_3\bar{Y}\epsilon_0 + g_4\left(1 + \frac{\epsilon_1}{2} - \frac{5\epsilon_1^2}{8}\right) \tag{4.14}$$

Squaring on both sides we get

$$(P_2 - \bar{Y})^2 = \bar{Y}^2 + \bar{Y}^2 g_3^2(1+\epsilon_0^2) + g_4^2(1 - \epsilon_1^2) - 2g_3\bar{Y}^2 - 2g_4\bar{Y}\left(1 - \frac{5\epsilon_1^2}{8}\right) + 2g_3 g_4 \bar{Y}\left(1 - \frac{5\epsilon_1^2}{8} + \frac{\epsilon_0\epsilon_1}{2}\right) \tag{4.15}$$

Taking expectations up to the first order approximation, we get mean square error (MSE),

$$MSE(P_2) = \bar{Y}^2 V_y + g_3^2 A_2 + g_4^2 B_2 + 2g_3 C_2 + 2g_4 D_2 + 2g_3 g_4 E_2 \tag{4.16}$$

where,

$$A_2 = \bar{Y}^2(1 + V_y)$$

$$B_2 = 1 - V_x$$

$$C_2 = \bar{Y}^2$$

$$D_2 = \bar{Y}\left(1 - \frac{5}{8}V_x\right)$$

$$E_2 = \bar{Y}\left(1 - \frac{5}{8}V_x + \frac{1}{2}V_{yx}\right)$$

To find out the minimum MSE for $P_2$, we partially differentiate equation (4.16) wrt $g_3$ & $g_4$ and equating to zero we get

$$g_3^* = \frac{B_2 C_2 - D_2 E_2}{A_2 B_2 - E_2^2} \tag{4.17}$$

$$g_4^* = \frac{A_2 D_2 - C_2 E_2}{A_2 B_2 - E_2^2} \tag{4.18}$$

Putting the optimum value of $g_3$ & $g_4$ in the equation (4.16), we get a minimum MSE of $P_2$ as

$$\text{Min } MSE = C_2 + \frac{B_2 C_2^2 + A_2 D_2^2 - 2C_2 D_2 E_2}{E_2^2 - A_2 B_2} \tag{4.19}$$

3.) $P_3 = g_5 \bar{y}_{rss}^* + g_6\left(\frac{\bar{X}}{\bar{x}_{rss}^*}\right) exp\left(\frac{\bar{X} - \bar{x}_{rss}^*}{\bar{X} + \bar{x}_{rss}^*}\right)$ \hfill (4.20)

Expressing $P_3$ given in equation (4.20) in terms of $\epsilon$'s we get

$$P_3 = g_5\bar{Y}(1+\epsilon_0) + g_6(1+\epsilon_1)^{-1} \exp\left(\frac{-\epsilon_1}{2+\epsilon_1}\right) \tag{4.21}$$

$$P_3 - \bar{Y} = (g_5 - 1)\bar{Y} + g_5\bar{Y}\epsilon_0 + g_6\left(1 - \frac{3\epsilon_1}{2} + \frac{15\epsilon_1^2}{8}\right) \tag{4.22}$$

$$\text{Bias }(P_3) = \bar{Y}(g_5 - 1) + g_6\left[1 + \frac{15}{8}V_x\right] \tag{4.23}$$

**CASE 1: SUM OF WEIGHTS IS UNITY ($g_5 + g_6 = 1$)**

The MSE of the estimator $P_3$ is shown as

$$MSE(P_3) = \bar{Y}^2[V_y + g_6^2 V_x - 2g_6 V_{yx}] \tag{4.24}$$

To find out the minimum MSE for $P_3$, we partially differentiate equation (4.24) wrt and equating to zero we get

$$g_6^* = \frac{V_{yx}}{V_x} \tag{4.25}$$

Putting the optimum value of $g_6$ in the equation (4.24), we get a minimum MSE of $P_3$ as

$$\text{Min MSE} = \bar{Y}^2\left(V_y - \frac{V_{yx}^2}{V_x}\right) \tag{4.26}$$

**CASE 2: THE SUM OF WEIGHTS IS FLEXIBLE ($g_5 + g_6 \neq 1$)**

$$P_3 - \bar{Y} = (g_5 - 1)\bar{Y} + g_5\bar{Y}\epsilon_0 + g_6\left(1 - \frac{3\epsilon_1}{2} + \frac{15\epsilon_1^2}{8}\right) \tag{4.27}$$

Squaring on both sides we get

$$(P_3 - \bar{Y})^2 = \bar{Y}^2 + \bar{Y}^2 g_5^2(1+\epsilon_0^2) + g_6^2(1+6\epsilon_1^2) - 2g_5\bar{Y}^2 - 2g_6\bar{Y}\left(1 + \frac{15\epsilon_1^2}{8}\right)$$
$$+ 2g_5 g_6 \bar{Y}\left(1 + \frac{15\epsilon_1^2}{8} - \frac{3\epsilon_0\epsilon_1}{2}\right) \tag{4.28}$$

Taking expectations up to the first order approximation, we get mean square error (MSE),

$$MSE(P_3) = \bar{Y}^2 V_y + g_5^2 A_3 + g_6^2 B_3 + 2g_5 C_3 + 2g_6 D_3 + 2g_5 g_6 E_3 \tag{4.29}$$

where,

$$A_3 = \bar{Y}^2(1+V_y)$$
$$B_3 = 1 + 6V_x$$
$$C_3 = \bar{Y}^2$$
$$D_3 = \bar{Y}\left(1 + \frac{15}{8}V_x\right)$$

$$E_3 = \bar{Y}\left(1 + \frac{15}{8}V_x - \frac{3}{2}V_{yx}\right)$$

To find out the minimum MSE for $P_3$, we partially differentiate equation (4.29) wrt $g_5$ & $g_6$ and equating to zero we get

$$g_5^* = \frac{B_3 C_3 - D_3 E_3}{A_3 B_3 - E_3^2} \tag{4.30}$$

$$g_6^* = \frac{A_3 D_3 - C_3 E_3}{A_3 B_3 - E_3^2} \tag{4.31}$$

Putting the optimum value of $g_5$ & $g_6$ in the equation (4.29), we get a minimum MSE of $P_3$ as

$$\text{Min } MSE = C_3 + \frac{B_3 C_3^2 + A_3 D_3^2 - 2C_3 D_3 E_3}{E_3^2 - A_3 B_3} \tag{4.32}$$

## 5. Numerical Illustration

We assess the effectiveness of the recommended estimators with the other estimators taken into consideration in this paper in this section. We selected one real data set of the population in the case of positive correlation coefficient between Y and X in order to illustrate the characteristics of the recommended estimators. For the purpose of evaluating the qualities of the suggested estimators, the population data set is taken from Singh (2003). The data and parameter values are described in the sections below:

Y= true amount of non-real estate farm loans in different states during 1997, X=true amount of real estate farm loans in different states during 1997, $Y_{me}$ = observed amount of non-real estate farm loans in different states during 1997, and $X_{me}$ = observed amount of real estate farm loans in different states during 1997.

$N = 50, \mu_x = 170, \mu_y = 127, \sigma_X^2 = 3300, \sigma_y^2 = 1278, \rho_{xy} = 0.964, \sigma_v^2 = 36, \sigma_u^2 = 36,$
$N_1=30, N_2=20, \sigma_{X2}^2 = 3300, \sigma_{y2}^2 = 1278, \sigma_{v2}^2 = 3300, \sigma_{u2}^2 = 1278.$

Additionally, we generated two bivariate RSS samples from the population with N=50, one for the variables X, Y and the other for the error terms' variable U ,V with set size k = 3 and replication r = 4 where $r_1 = 3$ from response group and $r_2^{'} = 1$ from non-response group. The ranked set sampling technique described in Section 2 is used to draw the RSS sample concurrently for the true study and auxiliary variables and error terms. The formula for Percent Relative Efficiency (PRE), and percentage contribution of the measurement error (PCME) are defined, respectively, as

$$PRE(Estimators) = \frac{MSE(\bar{y}_{rss}^*)}{MSE(estimator)} \times 100 \tag{5.1}$$

$$PCME = \frac{MSE\,()_m - MSE\,()_0}{MSE\,()_0} \times 100 \tag{5.2}$$

where $MSE\ ()_0$ are the MSEs when there is no ME, and $MSE\ ()_m$ are the MSEs when there is ME.

**Table 1: The MSE, PRE and PCME of the Estimators**

| Estimators | $MSE\ ()_0$ | $MSE\ ()_m$ | PRE | PCME |
|---|---|---|---|---|
| $\bar{y}_{rss}^*$ | 282394.9 | 27935.58 | 100 | 2.108525 |
| $\bar{y}_{Re}$ | 262524.8 | 261745.1 | 107.5688 | 0.297885 |
| $\bar{y}_{De}$ | 251062.4 | 250693.2 | 112.4800 | 0.147267 |
| $\bar{y}_{exp}$ | 252285.6 | 252283.9 | 111.9346 | 0.000671 |
| $P_1$ | 131398.2 | 130584.4 | 214.9153 | 0.848049 |
| $P_2$ | 16345.8 | 14947.9 | 1727.6220 | 0.001602 |
| $P_3$ | 27935.5 | 26816.4 | 1010.8790 | 0.000315 |

## 6. Simulation Study

We perform some simulation experiments to check the recommended estimator's relative efficiency (RE) with the conventional, ratio, regression estimator and other existing estimators. This is done via the following steps

1. We have generated 4-variate random observations of size N=1000 from a 4-variate normal distribution with mean $(\mu_x, \mu_y, 0,0) = (170,125,0,0)$ and covariance matrix
$$\begin{bmatrix} \sigma_x^2 & \rho_{xy}\sigma_x\sigma_y & 0 & 0 \\ \rho_{xy}\sigma_x\sigma_y & \sigma_y^2 & 0 & 0 \\ 0 & 0 & \sigma_v^2 & \rho_{vu}\sigma_v\sigma_u \\ 0 & 0 & \rho_{vu}\sigma_v\sigma_u & \sigma_u^2 \end{bmatrix}$$, where we have $\sigma_x^2 = 3300$, $\sigma_y^2 = 1200$ and for error terms, we have $\sigma_v^2 = 36$, $\sigma_u^2 = 36$ and $\rho_{vu} = 0$.

2. The parameters were calculated for this simulated population of size N = 1000 with different level of non-response rate.

3. A sample of size n with $n_1$ and $n_2'$ has been selected for X, Y, U, V from this simulated population.

4. Use the sample data to obtain the MSE of all the estimators under study.

5. The entire process from step 3 to step 4 was replicated 10000 times to obtain MSEs, the average of the 10000 values obtained are the MSE of each estimator of population mean.

6. The formula has been used to determine the PRE of each estimator with regard to $\bar{y}_{rss}^*$.

**Table 2: The MSE, PRE and PCME of the Estimators for uncorrelated measurement errors for n=12, 15, 18 for $\rho_{xy}$ =0.9, 0.8, 0.7.**

| k=2 | Estimators | $\rho_{xy}$ =0.9 | | | $\rho_{xy}$ =0.8 | | | $\rho_{xy}$ =0.7 | | |
|---|---|---|---|---|---|---|---|---|---|---|
| $(r_1, r_2')$ | | MSE | PRE | PCME | MSE | PRE | PCME | MSE | PRE | PCME |
| (3,1) | $\bar{y}_{rss}^*$ | 93.52088 | 100 | 4.176303 | 98.58077 | 100 | 3.948439 | 102.9866 | 100 | 3.773128 |
| | $\bar{y}_{Re}$ | 36.75932 | 254 | 19.55712 | 62.7453 | 157 | 10.62688 | 88.19799 | 117 | 7.356067 |

| $(r_1, r_2')$ | Estimators | MSE | PRE | PCME | MSE | PRE | PCME | MSE | PRE | PCME |
|---|---|---|---|---|---|---|---|---|---|---|
| | $\bar{y}_{De}$ | 24.07662 | 388 | 26.14397 | 40.86887 | 241 | 13.27236 | 55.71295 | 185 | 9.023779 |
| | $\bar{y}_{exp}$ | 32.87364 | 284 | 15.10975 | 48.1688 | 205 | 9.840138 | 62.88749 | 164 | 7.374694 |
| | $P_1$ | 23.7056 | 395 | 26.35877 | 40.37013 | 244 | 13.29689 | 55.08652 | 187 | 9.017369 |
| | $P_2$ | 22.716 | 412 | 12.77613 | 26.91309 | 366 | 5.481219 | 28.33014 | 364 | 3.700778 |
| | $P_3$ | 9.84816 | 950 | 28.29456 | 17.86781 | 552 | 13.98362 | 25.66976 | 401 | 9.440866 |
| (3,2) | $\bar{y}_{rss}^*$ | 69.38075 | 100 | 4.408731 | 76.40575 | 100 | 3.986337 | 82.61588 | 100 | 3.68153 |
| | $\bar{y}_{Re}$ | 30.29328 | 229 | 18.48595 | 52.25157 | 146 | 9.981124 | 73.42529 | 113 | 6.933766 |
| | $\bar{y}_{De}$ | 21.13049 | 328 | 22.69789 | 36.493 | 209 | 11.44687 | 50.04298 | 165 | 7.773648 |
| | $\bar{y}_{exp}$ | 27.06366 | 256 | 14.2657 | 41.34061 | 185 | 8.910479 | 54.83378 | 151 | 6.590547 |
| | $P_1$ | 20.92152 | 332 | 22.78118 | 36.18307 | 211 | 11.44675 | 49.62903 | 166 | 7.760703 |
| | $P_2$ | 17.52592 | 396 | 9.947241 | 20.06768 | 381 | 4.613755 | 20.89053 | 395 | 3.319749 |
| | $P_3$ | 8.83233 | 786 | 24.00134 | 15.97343 | 478 | 12.07152 | 22.8113 | 362 | 8.230968 |
| (3,3) | $\bar{y}_{rss}^*$ | 56.57136 | 100 | 4.458211 | 62.73231 | 100 | 4.001619 | 68.18206 | 100 | 3.670632 |
| | $\bar{y}_{Re}$ | 25.10446 | 225 | 18.4095 | 43.26726 | 145 | 9.959277 | 60.77995 | 112 | 6.916132 |
| | $\bar{y}_{De}$ | 17.89674 | 316 | 21.95435 | 30.93561 | 203 | 11.08242 | 42.45659 | 161 | 7.50594 |
| | $\bar{y}_{exp}$ | 22.42608 | 252 | 14.18877 | 34.40429 | 182 | 8.825123 | 45.72228 | 149 | 6.50729 |
| | $P_1$ | 17.75732 | 319 | 22.0104 | 30.72542 | 204 | 11.07917 | 42.17324 | 162 | 7.4937 |
| | $P_2$ | 14.58792 | 388 | 9.139797 | 16.60814 | 378 | 4.282333 | 17.24761 | 395 | 3.126265 |
| | $P_3$ | 7.5389 | 750 | 23.08328 | 13.57847 | 462 | 11.70262 | 19.35188 | 352 | 7.990342 |
| k=3 | Estimators | $\rho_{xy}=0.9$ | | | $\rho_{xy}=0.8$ | | | $\rho_{xy}=0.7$ | | |
| $(r_1, r_2')$ | | MSE | PRE | PCME | MSE | PRE | PCME | MSE | PRE | PCME |
| (3,1) | $\bar{y}_{rss}^*$ | 131.2124 | 100 | 3.920209 | 135.4842 | 100 | 3.786941 | 138.9096 | 100 | 3.690026 |
| | $\bar{y}_{Re}$ | 47.26624 | 278 | 20.30226 | 80.3226 | 169 | 11.05518 | 112.9396 | 123 | 7.641181 |
| | $\bar{y}_{De}$ | 29.49531 | 445 | 29.15982 | 49.62112 | 273 | 14.84438 | 67.39613 | 206 | 10.11599 |
| | $\bar{y}_{exp}$ | 43.30854 | 303 | 15.17601 | 61.72708 | 219 | 10.18661 | 79.49242 | 175 | 7.741702 |
| | $P_1$ | 28.76973 | 456 | 29.63595 | 48.70487 | 278 | 14.92006 | 66.29274 | 210 | 10.12528 |
| | $P_2$ | 27.37173 | 479 | 15.26783 | 35.33503 | 383 | 5.938995 | 37.83615 | 367 | 4.638612 |
| | $P_3$ | 11.45074 | 1146 | 33.30097 | 21.08479 | 643 | 15.92393 | 30.50483 | 455 | 10.64441 |
| (3,2) | $\bar{y}_{rss}^*$ | 91.61935 | 100 | 4.279947 | 100.4154 | 100 | 3.890333 | 108.1287 | 100 | 3.609155 |
| | $\bar{y}_{Re}$ | 38.86486 | 236 | 18.56927 | 67.04488 | 150 | 10.02426 | 94.24744 | 115 | 6.959951 |
| | $\bar{y}_{De}$ | 26.79166 | 342 | 23.24062 | 46.23939 | 217 | 11.73237 | 63.40422 | 171 | 7.973058 |
| | $\bar{y}_{exp}$ | 35.15412 | 261 | 14.09152 | 53.38376 | 188 | 8.865378 | 70.5933 | 153 | 6.579766 |
| | $P_1$ | 26.43026 | 347 | 23.36966 | 45.71096 | 220 | 11.73769 | 62.70311 | 172 | 7.959175 |
| | $P_2$ | 22.34817 | 410 | 11.0329 | 25.82353 | 389 | 5.035824 | 26.97953 | 401 | 3.571553 |
| | $P_3$ | 10.98054 | 834 | 24.93247 | 19.94264 | 504 | 12.45161 | 28.52266 | 379 | 8.478814 |
| (3,3) | $\bar{y}_{rss}^*$ | 72.63082 | 100 | 4.372172 | 80.37625 | 100 | 3.932539 | 87.19464 | 100 | 3.61422 |
| | $\bar{y}_{Re}$ | 31.6243 | 230 | 18.46974 | 54.51088 | 147 | 9.987894 | 76.59619 | 114 | 6.933279 |
| | $\bar{y}_{De}$ | 22.46306 | 323 | 22.20405 | 38.82949 | 207 | 11.21359 | 53.30955 | 164 | 7.594661 |
| | $\bar{y}_{exp}$ | 28.5487 | 254 | 14.03866 | 43.64602 | 184 | 8.768134 | 57.89853 | 151 | 6.47843 |
| | $P_1$ | 22.23511 | 327 | 22.28267 | 38.48811 | 209 | 11.21141 | 52.85058 | 165 | 7.580177 |
| | $P_2$ | 18.32635 | 396 | 9.656348 | 20.9575 | 384 | 4.501479 | 21.80774 | 400 | 3.260403 |
| | $P_3$ | 9.33857 | 778 | 23.53815 | 16.85877 | 477 | 11.88697 | 24.04633 | 363 | 8.10964 |
| k=4 | Estimators | $\rho_{xy}=0.9$ | | | $\rho_{xy}=0.8$ | | | $\rho_{xy}=0.7$ | | |
| $(r_1, r_2')$ | | MSE | PRE | PCME | MSE | PRE | PCME | MSE | PRE | PCME |
| (3,1) | $\bar{y}_{rss}^*$ | 168.9039 | 100 | 3.778953 | 172.3877 | 100 | 3.694806 | 174.8327 | 100 | 3.641143 |
| | $\bar{y}_{Re}$ | 57.77317 | 292 | 20.78127 | 97.8999 | 176 | 11.33144 | 137.6811 | 127 | 7.824622 |

|       | Estimator | | | | | | | | |
|-------|-----------|---|---|---|---|---|---|---|---|
|       | $\bar{y}_{De}$ | 34.61833 | 488 | 31.71071 | 57.81619 | 298 | 16.1847 | 78.28923 | 223 | 11.05567 |
|       | $\bar{y}_{exp}$ | 53.74344 | 314 | 15.21655 | 75.28535 | 229 | 10.40943 | 96.09735 | 182 | 7.983226 |
|       | $P_1$ | 33.41678 | 505 | 32.54457 | 56.35701 | 306 | 16.33343 | 76.57944 | 228 | 11.09002 |
|       | $P_2$ | 33.19011 | 509 | 25.24825 | 41.73177 | 413 | 7.352802 | 45.9945 | 380 | 5.074902 |
|       | $P_3$ | 12.65465 | 1335 | 38.8305 | 23.8023 | 724 | 17.77877 | 34.73775 | 503 | 11.7402 |
| (3,2) | $\bar{y}_{rss}^*$ | 113.8579 | 100 | 4.201618 | 124.4251 | 100 | 3.831475 | 133.6415 | 100 | 3.564456 |
|       | $\bar{y}_{Re}$ | 47.43645 | 240 | 18.62259 | 81.83819 | 152 | 10.05182 | 115.0696 | 116 | 6.976666 |
|       | $\bar{y}_{De}$ | 32.34968 | 352 | 23.70436 | 55.78896 | 223 | 11.97982 | 76.48568 | 175 | 8.148485 |
|       | $\bar{y}_{exp}$ | 43.24457 | 263 | 13.98276 | 65.42691 | 190 | 8.836899 | 86.35283 | 155 | 6.572934 |
|       | $P_1$ | 31.79406 | 358 | 23.88443 | 54.98477 | 226 | 11.99203 | 75.42368 | 177 | 8.13437 |
|       | $P_2$ | 26.90833 | 423 | 12.289 | 31.36639 | 397 | 5.449903 | 32.87165 | 407 | 3.821724 |
|       | $P_3$ | 13.00797 | 875 | 25.81641 | 23.74224 | 524 | 12.79065 | 34.01609 | 393 | 8.694689 |
| (3,3) | $\bar{y}_{rss}^*$ | 88.69029 | 100 | 4.317365 | 98.02018 | 100 | 3.888366 | 106.2072 | 100 | 3.578047 |
|       | $\bar{y}_{Re}$ | 38.14414 | 233 | 18.50945 | 65.7545 | 149 | 10.00673 | 92.41243 | 115 | 6.944572 |
|       | $\bar{y}_{De}$ | 26.9829 | 329 | 22.42584 | 46.63608 | 210 | 11.33172 | 64.03939 | 166 | 7.676333 |
|       | $\bar{y}_{exp}$ | 34.67132 | 256 | 13.94178 | 52.88775 | 185 | 8.731116 | 70.07477 | 152 | 6.459577 |
|       | $P_1$ | 26.64476 | 333 | 22.52864 | 46.13222 | 212 | 11.33109 | 63.36333 | 168 | 7.659825 |
|       | $P_2$ | 21.97493 | 404 | 10.14075 | 25.22841 | 389 | 4.700793 | 26.29388 | 404 | 3.383227 |
|       | $P_3$ | 11.08064 | 800 | 23.96879 | 20.05747 | 489 | 12.05228 | 28.63578 | 371 | 8.213946 |

**Table 3: The MSE and PRE of the Estimators for different level of measurement errors.**

| Estimators | $\delta$ =5% | | $\delta$ =10% | | $\delta$ =15% | | $\delta$ =20% | | $\delta$ =25% | | $\delta$ =30% | |
|------------|-----|-----|-----|-----|-----|-----|-----|-----|-----|-----|-----|-----|
|            | MSE | PRE | MSE | PRE | MSE | PRE | MSE | PRE | MSE | PRE | MSE | PRE |
| $\bar{y}_{rss}^*$ | 285.1053 | 100 | 303.8445 | 100 | 330.6102 | 100 | 349.7005 | 100 | 368.7909 | 100 | 387.8812 | 100 |
| $\bar{y}_{Re}$ | 106.8639 | 267 | 145.7258 | 209 | 185.9683 | 178 | 224.1228 | 156 | 262.2774 | 141 | 300.4319 | 129 |
| $\bar{y}_{De}$ | 90.86182 | 314 | 121.7033 | 250 | 154.9283 | 213 | 182.9231 | 191 | 209.8797 | 176 | 235.9944 | 164 |
| $\bar{y}_{exp}$ | 128.3303 | 222 | 152.1002 | 200 | 181.019 | 183 | 204.8754 | 171 | 228.7318 | 161 | 252.5882 | 154 |
| $P_1$ | 88.72115 | 321 | 118.9056 | 256 | 151.2884 | 219 | 178.5023 | 196 | 204.6299 | 180 | 229.8695 | 169 |
| $P_2$ | 45.42354 | 628 | 55.88878 | 544 | 63.53246 | 520 | 70.78837 | 494 | 77.30125 | 477 | 83.28185 | 466 |
| $P_3$ | 32.40426 | 880 | 45.50265 | 668 | 59.24853 | 558 | 71.81682 | 487 | 84.22728 | 438 | 96.49494 | 402 |

**Discussion**

MSEs, PREs and PCMEs of the existing and recommended estimators using RSS are given when there is proximity of uncorrelated ME and NRE. It is evident from the table that proposed estimators have performed better (lesser MSE and greater PRE) over existing estimators and $P_2$ has proven to be superior to all other estimators. See PCME values for the outcome of MEs.

Table 2 shows the MSEs, PREs, and PCMEs of the existing and recommended estimators employing RSS when there are proximity of errors (ME and NRE) for n=12,15,18 and $\rho_{xy}$= 0.7,0.8,0.9 for k=2,3,4. The increase in sample size decreases the MSEs for all estimators. As the $\rho_{xy}$ increases, the MSE decreases for all estimators. The MSE of the estimators rises as the non-response rate rises. Additionally, it has been found that the PRE rises when Y and X's correlation coefficient increases. The PRE also increases when the total sample size n increases, but it lowers as the non-response rate k is elevated. We see proposed estimators have performed better over existing estimators and $P_3$ has shown supremacy over all others estimators.

Table 3 shows MSEs for different levels of measurement errors ($\delta$) for $\rho_{uv} = 0$. To get an idea about this, we presume that and $\delta = \frac{\sigma_u^2}{\sigma_v^2} = \frac{\sigma_y^2}{\sigma_x^2}$ and that the ratio of ME variance to real variance is the same. The values of MSE under $\sigma_u^2 > 0$, $\sigma_v^2 > 0$ are higher than the values of MSE under $\sigma_u^2 = \sigma_u^2 = 0$. As the magnitude of measurement errors rises, MSEs rise as well. This demonstrates conclusively that measurement errors cause the estimators' MSE values to go up.

From Table 2 and Table 3 we can say that the presence of errors (ME and NRE) does affect the statistical properties of estimators.

**Conclusion**

By utilizing auxiliary information, we have proposed RSS estimators for the population mean in the presence of errors (ME and NRE) on both Y and X. The bias and MSE of the proposed estimators were calculated up-to first order approximation. The recommended estimators were compared to existing estimator by using one natural population and one simulated population. Through numerical illustrations and simulated studies, we discovered that the suggested estimators outperformed existing estimators and $P_3$ has shown supremacy over all other estimators.

The simulation findings make it abundantly evident that errors (ME and NRE) affect characteristics of the estimators. Through simulation and numerical illustrations, we determined the PCME values of the recommended estimators for the effect of measurement errors. We discover that appropriate safety measures are needed to handle the excessive PCME values.

Based on our empirical study and simulation studies, we can conclude that our proposed estimators can be preferred over the other estimators taken in this paper in several real situations like agriculture sciences, mathematical sciences, biological sciences, poultry, business, economics, commerce, social sciences, etc.

Since there aren't any RSS estimators in the existence of errors (ME and NRE), more research can be conducted in a variety of methods, including by dynamic estimators. Other RSS methods, such as median RSS, double RSS, quartile RSS, extreme RSS, unbalanced RSS, and so forth, can be used in place of RSS to examine the effects of errors (ME and NRE).